\documentstyle[11pt,newpasp,twoside,epsf]{article}
\newcommand{\hst}{{\it HST\,\,}}

\def\edcomment#1{\iffalse\marginpar{\raggedright\sl#1\/}\else\relax\fi}
\reversemarginpar

\begin{document}
\title{Morphology of AGN in the central kiloparsec}
\author{Paul Martini}
\affil{Observatories of the Carnegie Institution of Washington, 813 Santa 
Barbara St., Pasadena, CA 91101, USA}

\begin{abstract}
Hubble Space Telescope observations of the central kiloparsec of AGN have
revealed a wealth of structure, particularly nuclear bars and spirals,
that are distinct from analogous features in the disks of spiral galaxies.
WFPC2 and NICMOS images of a large sample of AGN observed at high spatial
resolution make it possible to quantify the frequency and detailed
properties of these structures. Nearly all AGN have nuclear spiral dust
lanes in the central kiloparsec, while only a small minority contain
nuclear bars. If these nuclear dust spirals trace shocks in the 
circumnuclear, gaseous disks, they may dissipate sufficient angular
momentum to fuel the active nucleus.
\end{abstract}

\section{Introduction}

Nearly all galaxies appear to have central, supermassive black holes, yet 
only a small fraction host AGN. To fuel the mass accretion rates of $\sim 0.01
M_{\odot}$ yr$^{-1}$ that powers nuclear activity in the Seyfert galaxies 
discussed here, there must be mechanisms that remove angular momentum from the 
host galaxy ISM and drive it within range of the central black hole. 
Galaxy interactions and bars can remove angular momentum and fuel nuclear 
activity; however, there remain many AGN that show no evidence of either a bar 
or a recent interaction. 

We undertook a survey of Seyfert 2s from the CfA redshift survey with \hst 
to look for evidence of additional mechanisms, such as nuclear bars 
(Shlosman et al. 1989; Pfenniger \& Norman 1990), that could transport the 
host galaxy ISM from 
100-parsec scales into the active nucleus. If fuel is continuously transported 
inwards, then evidence for gas flow traced by the dust morphology should 
be present. We have analyzed visible and near-infrared images from \hst to 
map the stellar surface density and dust distribution in 24 Seyfert 2s and 
found a small percentage of them host nuclear bars, but nearly all have 
nuclear dust spirals. 

\section{Looking for nuclear bars}

Nuclear bars are intrinsically more difficult to detect and characterize 
than large-scale bars as their semimajor axis lengths are on order one 
fifth as large. As the nearby AGN population is generally more distant than 
comparable non-active galaxies, the frequency of nuclear bars in AGN could not 
be examined with large samples before \hst and NICMOS. To identify stellar 
nuclear bars we adopted the same selection criteria described by Mulchaey et 
al. (1997) and Knapen et al. (2000). These selection methods essentially look 
for increases in ellipticity at constant position angle, followed by a 
decrease in ellipticity at the end of the bar. Using these criteria we 
found evidence for stellar nuclear bars in the four galaxies shown in 
Figure~1: Mrk~573, NGC~5283, Mrk~471, and NGC~5929 (Martini et al. 2001). 
Our detection rate of $\sim 25$\% (after accounting for selection effects) is 
similar to the detection rate reported by Erwin \& Sparke (1999) and 
M\'arquez et al.\ (2000) for non-active galaxies. The apparently equal and 
low percentage of nuclear bars in active and non-active galaxies suggests 
that nuclear bars are not responsible for fueling most Seyfert galaxies. 

\begin{figure}
\plotone{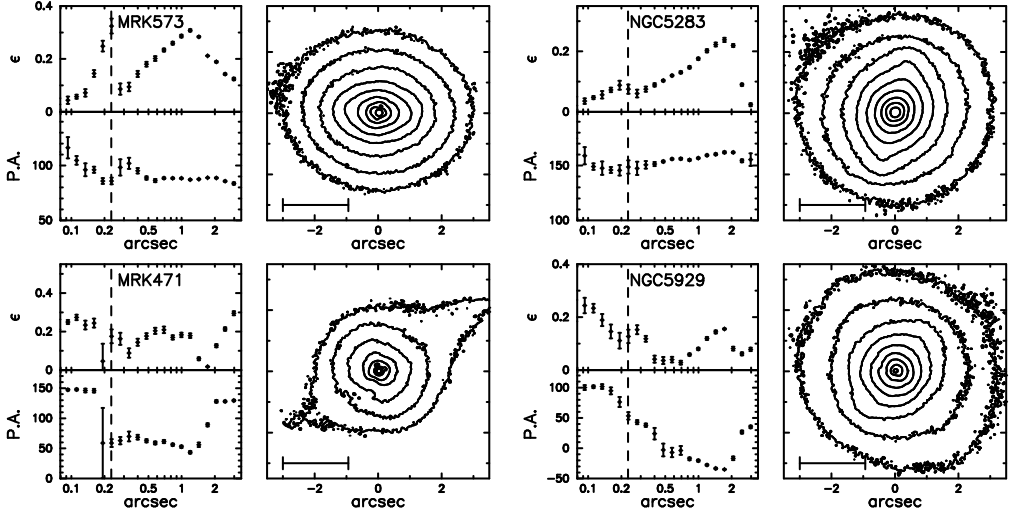}
\caption{The four stellar nuclear bar candidates identified in an isophotal 
analysis of NICMOS images of the CfA Seyfert 2s. For each galaxy the 
ellipticity and position angle distribution as a function of semimajor 
axis length is shown in the left panels and the F160W surface brightness 
contours in the right panels. The bar in the lower left corner of the 
contour plots corresponds to 1 kpc.}
\end{figure}

\begin{figure}
\plotone{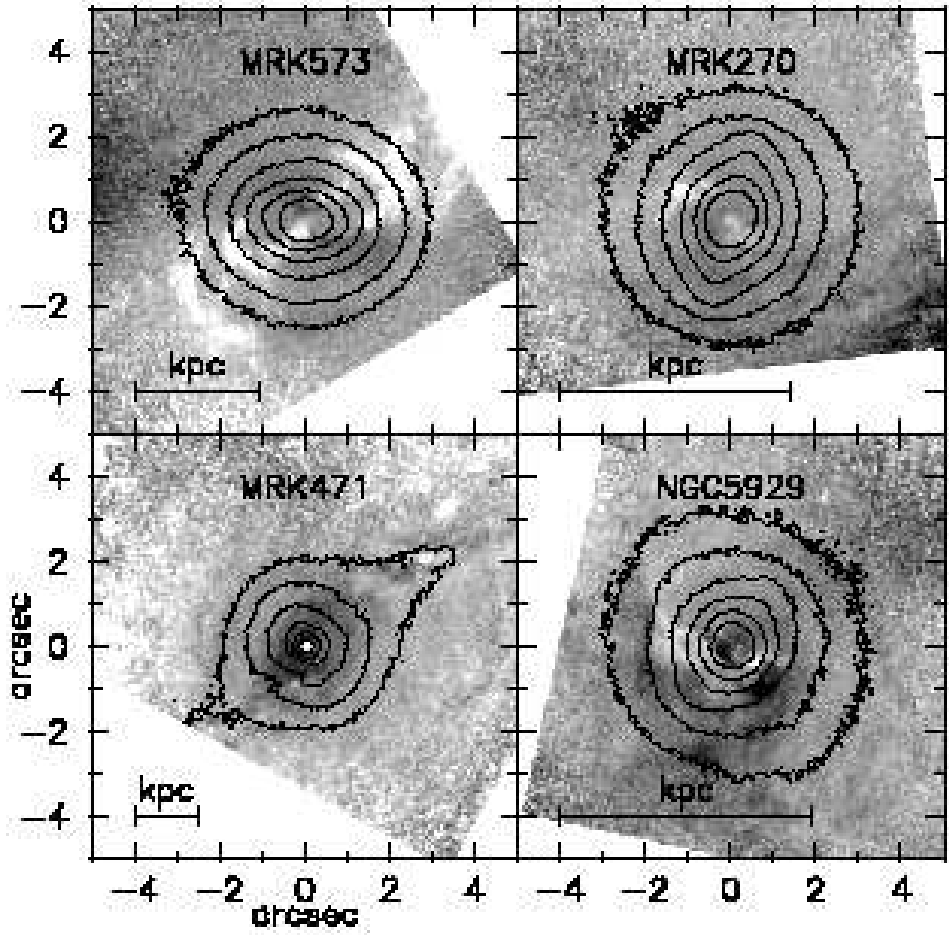}
\caption{$V-H$ color maps of the four galaxies with nuclear 
bar candidates from Martini \& Pogge (1999). The darker greyscale corresponds 
to greater reddening and traces the dust distribution. The contours are the 
same as in Figure~1.}
\end{figure}

The $V-H$ colormaps of Mrk~573, NGC~5283, and Mrk~471 shown in Figure~2 
show evidence for straight dust lanes crossing their nuclei, while none of 
the AGN without nuclear bars have similar straight dust lanes. This suggests 
that the gas flow traced by the dust is associated with the nuclear bar. 
Mrk~573 and NGC~5283 are particularly striking cases as the morphology of 
the dust lanes mimics the dust pattern seen in large-scale bars 
(Quillen et al. 1995; Regan et al 1997). The dust 
lanes appear to trace the leading edges of the nuclear stellar bar and then 
cut across the nucleus as a straight dust lane along the nuclear bar 
minor axis. 

\section{Nuclear spirals as a potential fueling mechanism}

While this study did not uncover a large percentage of nuclear bars in 
AGN, essentially all of these galaxies do have nuclear dust spirals on 
100-parsec scales (Martini \& Pogge 1999). These nuclear dust spirals can be 
morphologically classified as primarily ``flocculent'' or ``grand-design''  
in appearance, analogous to the spiral structure on kiloparsec scales in 
spiral galaxies. Examples of grand-design nuclear spirals in 
AGN are shown in Figure~3 ({\it top row}) and examples of flocculent spiral 
structure are shown in Figure~4 ({\it top row}). 

\begin{figure}
\plotone{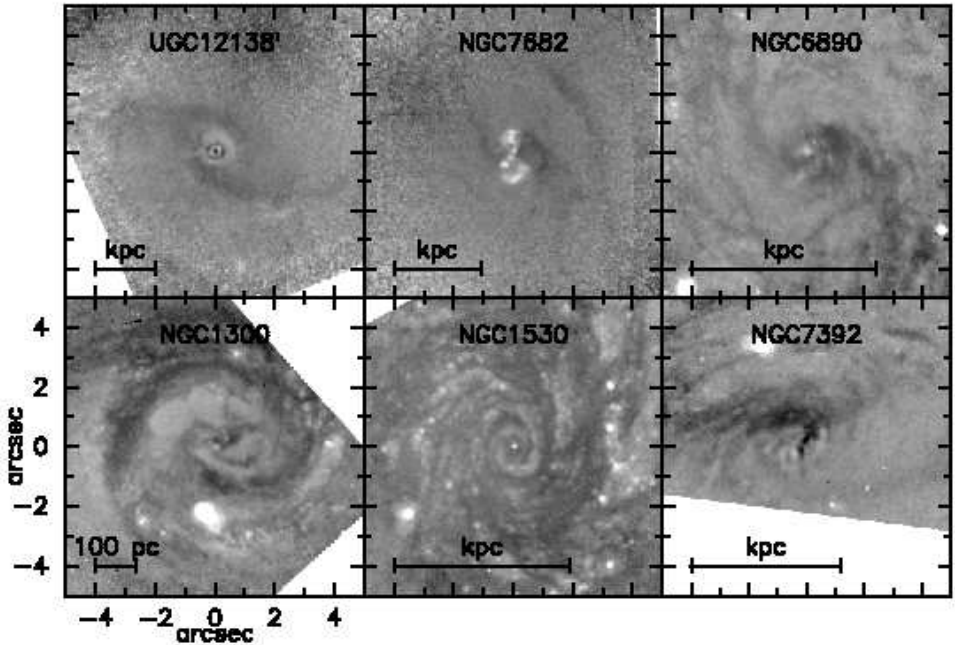}
\caption{$V-H$ colormaps of galaxies with grand-design nuclear spiral 
structure.}
\end{figure}

Gaseous disks in the central kiloparsec of spiral galaxies are generally not 
self-gravitating and are in nearly solid-body rotation, in contrast to their 
main disks on kiloparsec scales, which are prone to gravitational 
instabilities and subject to considerable differential rotation. Instead of 
gravity, pressure dominates the gas response to perturbations in these 
central, gaseous disks and the presence of nuclear dust spirals on 
these scales are evidence for gas flow driven by hydrodynamic effects. 
Englmaier \& Shlosman (2000) proposed that gas inflow from a large-scale 
bar can create grand-design nuclear spiral structure, such as that shown in 
Figure~3. All of the galaxies in our sample which show this grand-design 
spiral structure do have large-scale bars, which supports this model for the 
dust morphology. The flocculent spiral structure shown in Figure~4 may be 
due to the acoustic instabilities described by Elmegreen et al. (1998). 
In this model, small instabilities propagate as pressure-driven waves and are 
sheared by the small differential rotation in these disks to produce the 
flocculent structure (Montenegro et al. 1999). 
Both of these models for nuclear spiral structure propose that shocks create 
the density enhancements traced by these dust lanes. Shocks in these nearly 
ubiquitous spiral dust lanes is thus interesting in the context of AGN fueling 
as these shocks may dissipate sufficient angular momentum to fuel 
nuclear activity. Estimates of the rate of mass inflow in the context of 
theoretical models for both types of nuclear spiral structure are clearly 
needed to better assess its viability as a fueling mechanism for AGN. 

\begin{figure}
\plotone{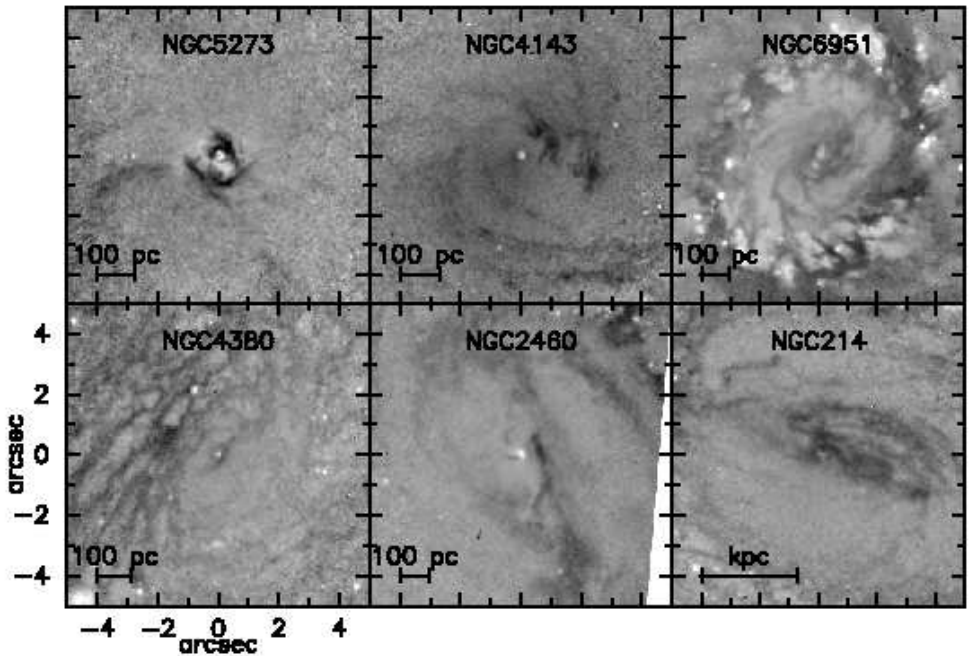}
\caption{Same as Figure~3 for galaxies with flocculent nuclear spiral 
structure.}
\end{figure}

Nuclear spiral structure has long been know to exist in nearby, non-active 
galaxies such as M100 and M101 (Sandage 1961), although the frequency of this 
structure has not been well characterized as it is rarely resolved in 
ground-based observations. Visible-wavelength \hst images of 
the centers of spirals have uncovered more non-active galaxies with nuclear 
spiral structure, but it is not obviously present in all cases (Carollo et al. 
1998). The near-ubiquity of nuclear dust spirals in AGN (see also Quillen et 
al 1999; Regan \& Mulchaey 1999) and the possibility that shocks remove 
sufficient angular momentum to fuel the nuclear activity prompts a simple 
question: Are nuclear dust spirals present with equal frequency in non-active 
galaxies? If they are not, this would be strong evidence in favor of a causal 
connection between nuclear dust spirals and the fueling of AGN. 

We are currently imaging a large sample of non-active galaxies with 
WFPC2 on \hst that have previous NICMOS images in order measure the frequency 
of nuclear spiral structure in a 
well-matched control sample. While these observations are not yet complete, 
we have found non-active galaxies which show grand-design 
(Figure~3, {\it bottom row}) and flocculent (Figure~4, {\it bottom row}) 
nuclear dust spirals. A preliminary result of this survey is that nuclear 
spirals structure is fairly common in non-active galaxies, although 
we do not yet have an adequate sample size to properly characterize the 
relative frequency of these structures in active and non-active galaxies, or 
to look for trends with Hubble type and other galaxy properties. 

\section{Implications for AGN fueling}

The main problem with nuclear spiral structure as a fueling mechanism for 
AGN is quite different from that of bars and interactions, namely it 
appears to be too common rather than too rare. These observations indicate 
that AGN and non-active galaxies have similar morphologies, whether nuclear 
spirals or nuclear bars, on 100-parsec scales. One simple interpretation is 
that the typical lifetime of low-luminosity AGN is less than the dynamical 
time on 100-parsec scales, which corresponds to a few times $10^7$ years. 

\acknowledgments

I would like to thank my collaborators in this project: Rick Pogge, John 
Mulchaey, and Mike Regan, for allowing me to present this work in advance of 
publication, as well as Johan Knapen for organizing such an interesting 
meeting. Support for this work was provided by NASA through grant numbers  
GO-7867 and GO-8597 from the Space Telescope Science Institute, which is
operated by the Association of Universities for Research in Astronomy,
Inc., under NASA contract NAS5-26555.


\begin{references}

\reference Carollo, C.M., Stiavelli, M., \& Mack, J. 1998, \aj, 116, 68
\reference Elmegreen, B.G., Elmegreen, D.M., Brinks, E., Yuan, C., Kaufman, M., Klari\'c, M., Montenegro, L., Struck, C., \& Thomasson, M. 1998, \apj, 503, L119
\reference Friedli, D. \& Martinet, L. 1993, \aap, 277, 27
\reference Knapen, J.H., Shlosman, I., \& Peletier, R.F. 2000, \apj, 529, 93
\reference M\'arquez, I. et al. 1999, \aaps, 140, 1
\reference Martini, P. \& Pogge, R.W. 1999, \aj, 118, 2646
\reference Martini, P., Pogge, R.W., Ravindranath, S., \& An, J.H. 2001, \apjs, {\it submitted}
\reference Montenegro, L.E., Yuan, C., \& Elmegreen, B.G. 1999, \apj, 520, 582
\reference Mulchaey, J.S., Regan, M.W., \& Kundu, A. 1997, \apjs, 11, 299
\reference Pfenniger, D. \& Norman, C. 1990, \apj, 363, 391
\reference Quillen, A.C., Frogel, J.A., Kenney, J.D.P., Pogge, R.W., \& 
DePoy, D.L. 1995, \apj, 441, 549
\reference Quillen, A.C., Alonso-Herrero, A., Rieke, M.J., McDonald, C., 
Falcke, H., \& Rieke, G.H. 1999, \apj, 525, 685
\reference Regan, M.W., Vogel, S.N., \& Teuben, P.J. 1997, \apj, 482, L143
\reference Regan, M.W. \& Mulchaey, J.S. 1999, \aj, 117, 2676
\reference Sandage, A. 1961, The Hubble Atlas of Galaxies. Carnegie 
Institution of Washington, Washington DC
\reference Shlosman, I., Frank, J., \& Begelman, M.C. 1989, Nature, 338, 45

\end{references}
\end{document}